\documentclass[11pt]{article}

\usepackage[preprint]{acl}
\usepackage{times}
\usepackage{tabularx}
\usepackage{siunitx}
\usepackage[utf8]{inputenc}
\usepackage[T1]{fontenc}
\usepackage{hyperref}
\usepackage{url}
\hypersetup{
    colorlinks=true,
    linkcolor=red,
    citecolor=teal,
    urlcolor=cyan,
}
\usepackage{booktabs}
\usepackage{amsfonts}
\usepackage{nicefrac}
\usepackage{microtype}
\usepackage{xcolor}
\usepackage{makecell}
\usepackage{amsthm}
\usepackage{subcaption}
\usepackage[export]{adjustbox}
\usepackage{threeparttable} 
\usepackage{latexsym}
\usepackage{bm}
\usepackage{xspace}
\usepackage[namelimits]{amsmath}
\usepackage{algcompatible}
\usepackage{subcaption}
\usepackage{inconsolata}
\usepackage{algorithm}
\usepackage{colortbl}
\usepackage[noend]{algpseudocode}
\usepackage{array} 
\usepackage{etoc}
\usepackage{graphicx}
\usepackage{multirow}
\usepackage{longtable}
\usepackage{tcolorbox}
\usepackage{lipsum}
\usepackage{footnote}  
\usepackage{natbib}
\usepackage{multirow}
\usepackage{enumitem}
\usepackage{listings}
\etocdepthtag.toc{mtchapter}
\etocsettagdepth{mtchapter}{subsubsection}
\etocsettagdepth{mtappendix}{none}
\usepackage{wrapfig}
\usepackage{arydshln}
\usepackage{threeparttable}
\usepackage{pdfpages}

\newcommand{\methodname}{EvidFuse}

\title{EvidFuse: Writing-Time Evidence Learning \\for Consistent Text–Chart Data Reporting}

\author{
 \textbf{Huanxiang Lin\textsuperscript{1}\thanks{Equal contribution: \href{sehuanxianglin@scut.mail.edu.cn}{sehuanxianglin@scut.mail.edu.cn}, \href{202420145083@scut.mail.edu.cn}{202420145083@scut.mail.edu.cn}}},
 \textbf{Qianyue Wang\textsuperscript{1,2}\footnotemark[1]},
 \textbf{Jinwu Hu\textsuperscript{1,2}},
 \textbf{Bailin Chen\textsuperscript{1,2}},
\\
 \textbf{Qing Du\textsuperscript{1,2}\thanks{Corresponding author: \href{duqing@scut.edu.cn}{duqing@scut.edu.cn}, \href{mingkuitan@scut.edu.cn}{mingkuitan@scut.edu.cn}}},
 \textbf{Mingkui Tan\textsuperscript{1,2}\footnotemark[2]},
\\
\\
 \textsuperscript{1}South China University of Technology,
 \textsuperscript{2}Pazhou Laboratory,
}

\begin{document}
\maketitle
\begin{abstract}
Data-driven reports communicate decision-relevant insights by tightly interleaving narrative text with charts grounded in underlying tables. However, current LLM-based systems typically generate narratives and visualizations in staged pipelines, following either a text-first-graph-second or a graph-first-text-second paradigm. These designs often lead to chart-text inconsistency and insight freezing, where the intermediate evidence space becomes fixed and the model can no longer retrieve or construct new visual evidence as the narrative evolves, resulting in shallow and predefined analysis. To address the limitations, we propose \textbf{\methodname{}}, a training-free multi-agent framework that enables writing-time text-chart interleaved generation for data-driven reports. \methodname{} decouples visualization analysis from long-form drafting via two collaborating components: a \textbf{Data-Augmented Analysis Agent}, equipped with Exploratory Data Analysis (EDA)-derived knowledge and access to raw tables, and a \textbf{Real-Time Evidence Construction Writer} that plans an outline and drafts the report while intermittently issuing fine-grained analysis requests. This design allows visual evidence to be constructed and incorporated exactly when the narrative requires it, directly constraining subsequent claims and enabling on-demand expansion of the evidence space. Experiments demonstrate that \methodname{} attains the top rank in both LLM-as-a-judge and human evaluations on chart quality, chart-text alignment, and report-level usefulness.
\end{abstract}

\section{Introduction}

Data-driven reports are a primary medium for communicating complex datasets in decision-making scenarios, ranging from public policy~\cite{policy} and scientific analysis~\cite{sci} to business intelligence~\cite{bi}. However, manually writing such reports is costly and time-consuming since analysts must repeatedly explore data, design appropriate visualizations, and craft a coherent narrative, while the quality and style can vary substantially across writers~\cite{human_writer}. With the rapid progress of large language models in long-form generation~\cite{re3,dome}, automating data-driven report writing has become increasingly feasible~\cite{FinRpt, Multi-Agent_Collaboration_for_Investment_Guidance, MEIT}.

\begin{figure}[!t]
    \centering
    \includegraphics[width=0.48\textwidth]{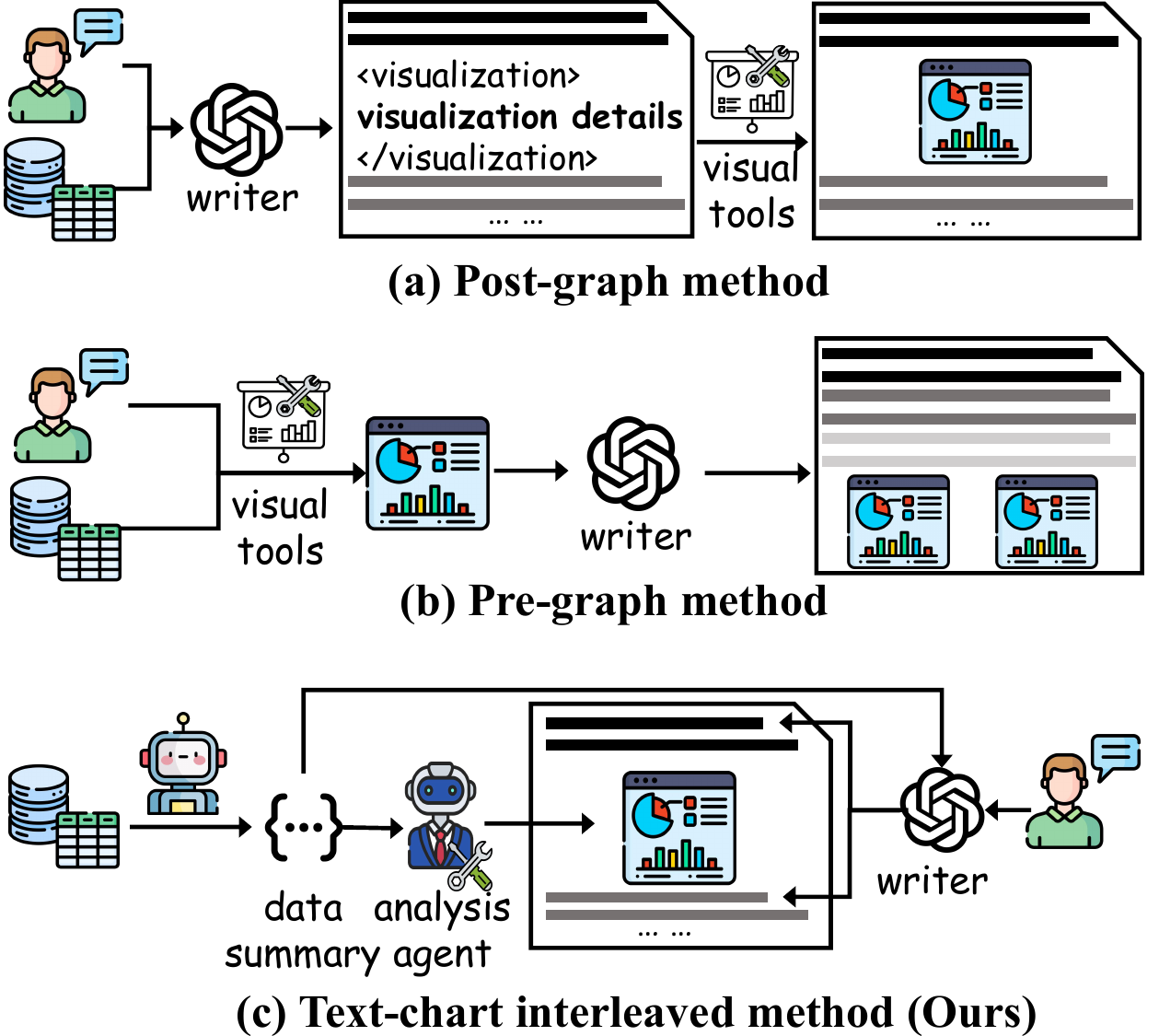}
    \caption{The illustration of different generation paradigms for text-chart interleaved report.}
    \label{fig: compare}
\end{figure}

\textit{Unfortunately}, generating data-driven reports with LLMs remains challenging. Such reports are typically produced in response to a concrete analysis request over multiple, dispersed tables and presented in a tightly text--chart interleaved form, where narrative claims and visualizations serve as mutually reinforcing evidence~\cite{yang2025multimodaldeepresearchergeneratingtextchart}. Two key challenges arise:
\textbf{1) On-demand Grounded Visualization:} The report requires retrieving and constructing the right chart as grounded evidence from multiple data tables at the exact point the narrative needs it, ensuring strict chart–text consistency~\cite{yang2025multimodaldeepresearchergeneratingtextchart}. \textbf{2) Decision-oriented Insight Depth:} The report requires composing decision-oriented, multi-step insights that go beyond surface description of visualization~\cite{islam-etal-2024-datanarrative, goal_driven_data_story}.

Most existing methods to text-chart report generation adopt \emph{staged} pipelines, which can be broadly categorized into two paradigms, as shown in Figure~\ref{fig: compare}. 
\textbf{Text-first-graph-second} methods first draft the narrative under a fixed set of inferred insights and the user request, and only afterward invoke external tools to create visualizations from chart specifications embedded in the generated text~\cite{islam-etal-2024-datanarrative,yang2025multimodaldeepresearchergeneratingtextchart}. 
Because the narrative is written without access to the actual visual evidence, these methods often produce claims that are weakly grounded or inconsistent with the resulting charts. 
\textbf{Graph-first-text-second} methods precompute a set of candidate figures from the user request and then generate text conditioned on these figures~\cite{dibia-2023-lida,deepanalyze}. 
While this improves visualization customization, the narrative often lacks explicit, verifiable correspondence between specific claims and specific figures, which undermines text-chart alignment. 
More fundamentally, both paradigms freeze the intermediate evidence space: once a set of insights or figures is fixed, the model cannot retrieve or construct new visual evidence as the narrative evolves. 
Therefore, existing systems tend to produce surface-level descriptions or predefined insights, falling short for  decision-oriented analysis.

To address the above limitations, we propose~\methodname{}, a training-free multi-agent framework for writing-time text--chart interleaved generation in data-driven reports. 
Specifically, \methodname{} comprises two collaborating components that decouple visualization analysis from long-form drafting, namely a \textbf{Data-Augmented Analysis Agent} and a \textbf{Real-Time Evidence Construction Writer}. 
The Data-Augmented Analysis Agent is equipped with Exploratory Data Analysis (EDA)-derived knowledge and dataset-specific background, including dataset summaries and access to raw tables. It responds to fine-grained analysis requests by constructing grounded visualizations together with corresponding captions on demand. 
The Real-Time Evidence Construction Writer is built on a multimodal LLM. It first plans an outline and then drafts the report while intermittently emitting analysis requests. Whenever a request is issued, generation is paused and the returned visual evidence is injected into the context to constrain subsequent claims. Drafting then resumes until \texttt{<EOS>} is produced. 
With this writing-time evidence construction mechanism, \methodname{} inserts visual evidence exactly when the narrative requires it. This design improves chart--text consistency and enables deeper, decision-relevant analysis without being restricted to a fixed set of precomputed insights. Our main contributions are as follows:

\begin{itemize}
    \item \textbf{A paradigm for text-chart interleaved generation of data reports.}
    We pinpoint text-chart inconsistency as a consequence of staged pipelines and further propose an interleaving paradigm that generates and inserts visualization evidence into context during contextual writing, so that the subsequent contextual claims are conditioned on grounded visualization, improving text-chart consistency.

    \item \textbf{A multi-agent collaboration framework that enables the construction of visual evidence at writing time.} We propose \methodname{}, a collaborative framework composed of a \emph{Data-Augmented Analysis Agent} and a \emph{Real-Time Evidence Construction Writer}. By registering the analysis agent as a callable tool, the writer issues fine-grained analysis requests \emph{during} generation and receive grounded charts and evidence that are injected back into context. This writing-time data interaction tightly couples insight construction with narrative generation, enabling progressive, decision-oriented analysis beyond predefined insights.

    \item \textbf{Both LLM-as-a-Judge and human evaluators prefer the reports generated by \methodname{} at multiple levels of quality.} Across three diverse report sources and six criteria in chart, chapter and report-level evaluation,~\methodname{} ranks best on average by both automated ranking and human ranking in most cases, demonstrating that writing-time text-chart interleaving produces reports with stronge text-chart consistency and in-depth insights than staged pipelines.
\end{itemize}

\begin{figure*}[t]
    \centering
    \includegraphics[width=\textwidth]{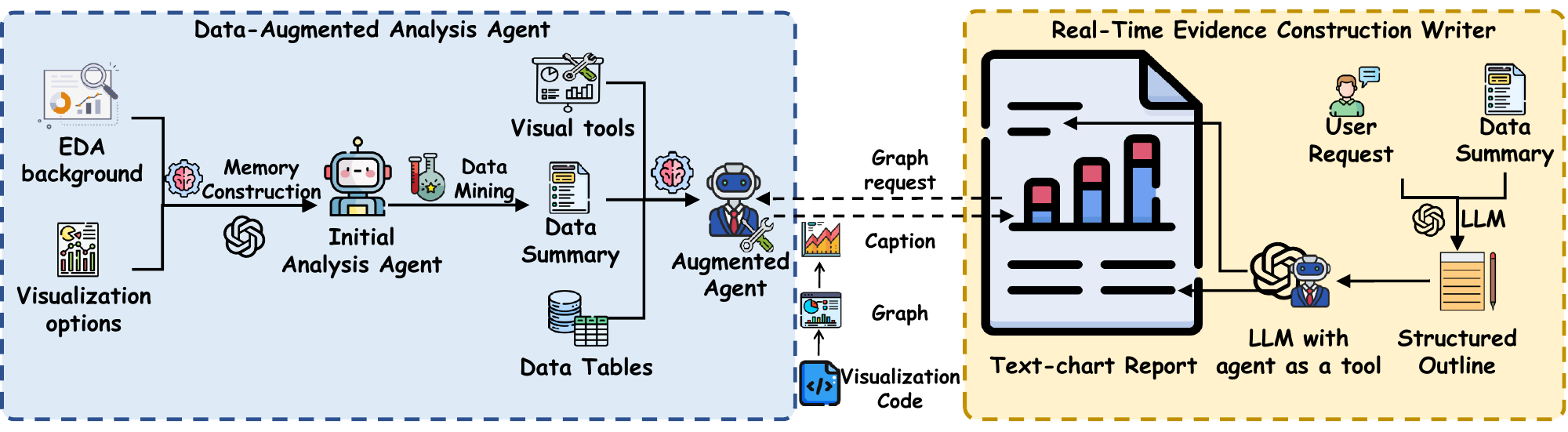}
    \caption{The Illustration of \methodname. Given a user analysis request and multiple data tables,~\methodname{} first specializes a \textbf{Data-Augmented Analysis Agent} by EDA-derived dataset overview and raw tables. A \textbf{Real-Time Evidence Construction Writer} then plans an outline and generates the report. During generation, the writer suspends by specific visualization requests and continues when it receives visualization and request-based captions from the analysis agent as context for subsequent generation until \texttt{<EOS>} terminates the generation.}
    \label{fig: method}
\end{figure*}

\section{Related Work}
\label{sec:related}

Automated data-driven report generation aims to synthesize analytical reports from structured datasets by tightly combining narrative text with data visualizations. Existing LLM-based approaches largely follow \emph{staged} generation pipelines, which can be categorized into two paradigms based on the generation order: \textbf{text-first--graph-second} and \textbf{graph-first--text-second}.

\textbf{Text-first--graph-second Paradigm.}
Methods in this paradigm first draft the narrative (often conditioned on extracted or inferred insights) and then insert visualizations post hoc, typically via placeholders or inline visualization instructions that are rendered afterward~\cite{yang2025multimodaldeepresearchergeneratingtextchart,islam-etal-2024-datanarrative}.
A representative method is DataNarrative~\cite{islam-etal-2024-datanarrative}, which generates a complete narrative and subsequently replaces the annotated chart positions with the visualization result. This paradigm benefits from long-form narration, but writing without the actual charts causes chart-text inconsistency once the narrative is finished.

\textbf{Graph-first--text-second Paradigm.}
Graph-first approaches precompute visual evidence, which is often a set of charts, from user intent and data, and then generate the report conditioned on these charts~\cite{dibia-2023-lida,deepanalyze,eda_insightpilot}.
DeepAnalyze~\cite{deepanalyze} exemplifies this direction by performing data analysis and visualization before report drafting.
Other systems for LLM-assisted data exploration and visualization, such as InsightPilot~\cite{eda_insightpilot} and LIDA~\cite{dibia-2023-lida}, naturally fit this paradigm as the graph-construction stage that translates analysis intent into chart specifications and provides the resulting charts as evidence for downstream narrative generation. This paradigm improves chart relevance and customization, but fixing the evidence space before writing can hinder iterative evidence seeking, limiting information depth in the report.

\section{Problem Setting}
We consider the problem of generating a data-driven analytical report from multiple data tables and the user analysis request. Given an input dataset \( D \) and a human request \( \text{Request} \), like task description, analysis goal, or reporting intent, our system produces a text-chart interleaved report.
\[
CT = \{t_1, v_1, \dots, t_n, v_n\},
\]
where each \( t_i \) is a textual segment and each \( v_i \) is a visual evidence in \( D \). The key challenge is to couple reliable data analysis with coherent narrative writing under the specific human request.

\section{Writing-Time Evidence Construction for Consistent Text–Chart Data Reports}

We propose~\methodname{}, a multi-agent collaboration framework for generating text-chart-interleaved reports driven by multiple data tables.
As shown in Figure~\ref{fig: method},~\methodname{} comprises a Data-Augmented Analysis Agent $A'$ and a Report Writer $W$ that enable write-time evidence construction through data interaction. Given data tables $D$ and a user request $R_{\text{user}}$, an initial MLLM-based analysis agent $A$ first constructs a data-table overview $DO$, which is then used to augment $A$ into $A'$ with access to the raw data tables.
The writer $W$ follows the plan-and-write paradigm to first generate an outline conditioned on $R_{\text{user}}$ and $DO$, and subsequently drafts the contextual content of the report. 
Whenever visual evidence is required to support a forthcoming claim, $W$ issues an analysis request delimited by \texttt{<visualization>} and \texttt{</visualization>}, and suspends generation.
In response, $A'$ queries the relevant data in $D$ and produces a grounded visualization, which is injected back into the context. $W$ then resumes writing conditioned on the returned visualization.
This interleaving loop repeats until $W$ generates \texttt{<EOS>}. 
By constructing visual evidence at writing time and conditioning subsequent text on the actual visualization,~\methodname{} improves text-chart consistency. The loop of evidence-write process enables dynamically deeper, decision-relevant insights tailored to $R_{\text{user}}$, rather than surface-level summaries.

\subsection{Data-Augmented Analysis Agent}

Generating visualization from multiple data tables on specific request is a non-trivial data analysis task, which requires selecting relevant variables, choosing appropriate plot types and aggregations, and producing interpretations~\cite{yang2025multimodaldeepresearchergeneratingtextchart}.
While coupling visualization and long-form report writing within a single model leads to redundant context and affects the task performance~\cite{cca}, we decouple writing from visualization by assigning data analysis and visualization workload to a dedicated Data-Augmented Analysis Agent.
The Data-Augmented Analysis Agent $A'$ is an MLLM-based backend equipped with tool-calling to a code-based visualization tool $T$ augmented with data specific background.
Given an analysis request $q$ and data tables $D$, $A'$ produces a visualization $v$ by $T$ with caption $g$, considering the visualization request. The detailed agent construction and visualization process is as follows.

\textbf{Agent Construction.} We initialize an MLLM-based analysis agent $A$ and update its memory with basic Exploratory Data Analysis (EDA) knowledge~\cite{eda}, including analysis concepts, statistical routines, visualization principles, and a set of visualization options.
Given a set of data tables $D=\{D_1,\ldots,D_m\}$, $A$ is then applied to construct a data overview $DO$ by conducting 5 to 8 EDA probe aspects tailored to the table types:
\begin{equation}
DO=A(P_{overview},D).
\label{eq: overview}
\end{equation}
This augmentation equips $A'$ with global context about the dataset via $DO$ and executable visualization capability via $T$, allowing it to ground requests in the underlying tables. $P_{\{\}}$ is task prompt which is the same as following prompt.

\textbf{Visualization Process.}
Given any analysis request $q$, $A'$ produces a visualization $v$ and an analysis-oriented caption $s$ through a systematic and iterative three-step procedure:

\textit{1) Visual specification planning.} $A'$ interprets the analysis request $q$ into an analysis intent $i$ and a comprehensive visualization specification $vs$ including referenced table subset $D_{ref}$ (see example in Appendix~\ref{sup:vis_spe}). This step is formatted as:
\begin{equation}
(i,vs)=A'(P_{ana},q).
\label{eq: vis_plan}
\end{equation}

\textit{2) Visualization via $T$.} Conditioned on the visualization specification $vs$, $T$ is called to generate executable code with an additional iterative refinement driven by visual-feedback (Details in Appendix~\ref{sup:vis_tool}). This step is formatted as:
\begin{equation}
c = T\!\left(D_{\text{ref}},\ vs;\ \mathrm{IterRefine}\right),
\label{eq: vis_tool}
\end{equation}
where $\mathrm{IterRefine}$ indicates an iterative loop that evaluates and updates the generated visualization based on textual feedback from MLLM.

\textit{3) Caption generation.}
$A'$ generates an analysis-oriented caption $s$ that directly responds to $q$ while strictly grounded in $c$. This step is formatted as:
\begin{equation}
s=A'(P_{caption}, c,i,q)
\label{eq: insight_for}
\end{equation}
The resulting pair $(c,s)$ is returned to the writer $W$ as comprehensive visual evidence $v$ to effectively condition subsequent report generation.

\subsection{Real-Time Evidence Construction Writer}

Existing methods typically follow a stage-wise pipeline, where visualization and narratives are produced asynchronously, resulting in text-chart inconsistency and limited insight depth in report~\cite{deepanalyze,islam-etal-2024-datanarrative}. To address these issues, we design a Real-Time Data-Interactive Writer $W$ that supports the construction of the visual evidence for textual content at write-time.

$W$ collaborates with the Data-Augmented Analysis Agent $A'$ to request, obtain and immediately incorporate visual evidence during contextual generation. Concretely, $W$ is responsible for outline planning and long-form narration, while delegating data interaction and visualization to $A'$. Whenever $W$ needs evidence to support a forthcoming claim, it emits an analysis request $q_i$ wrapped by \texttt{<visualization>} and \texttt{</visualization>} and suspends the generation process. $A'$ then returns a grounded visualization and caption as the context augmentation of writer. $W$ resumes writing conditioned on the returned visual evidence. This loop continue when \texttt{<EOS>} is generated.

By shifting from asynchronous, stage-wise generation to \emph{write-time evidence construction},~\methodname{} conditions narration on on-demand visual evidence, thereby improving text-chart consistency and enabling deeper, decision-oriented insights.

Concretely, the Report Writer $W$ is an MLLM-based agent. It follows the plan-write paradigm~\cite{Plan-and-Write,dome} to first be prompted to generate an outline conditioned on the user request and the data overview:
\begin{equation}
    O=W(P_{outline}, R_{user}, DO)
\label{eqa: outline}
\end{equation}

At each step, the writer produces a text segment including an analysis request limited by \texttt{<visualization>} and \texttt{</visualization>}:
\begin{equation}
x_i = \{t_i,\ q_i\} = W(P_{report}, H_{i-1}, O, DO),
\label{eqa: writer_output}
\end{equation}
where $t_i$ represents the generated textual content and $q_i$ is explicitly delimited by \texttt{<visualization>} and \texttt{</visualization>} tags, which triggers a suspension of the writer's generation. We then invoke the data-augmented analysis agent $A'$ to construct the corresponding visual evidence:

\begin{equation}
v_i=(c_i, s_i)=A'(q_i, D, DO;\ T),
\label{eqa: analysis_call}
\end{equation}
The history is updated by the generated content:
\begin{equation}
H_i = H_{i-1}\ \oplus\ v_i.
\label{eqa: history_update}
\end{equation}
$W$ resumes writing conditioned on the updated history $H_i$.
This interleaving process repeats until $W$ emits \texttt{<EOS>}. The final report $CT$ is the resulting history $H$ after termination. The complete procedure is formatted in Algorithm~\ref{alg: method}.

\begin{algorithm}[t]
\caption{Pipeline of \methodname{} for Text-Chart Interleaved Report Generation}
\label{alg: method}
\begin{algorithmic}[1]
\Require User request $R_{user}$, tables $D$, initial analysis agent $A$, visualization tool $T$, writer $W$, stop token set $Stop = \{\texttt{<EOS>}, \texttt{</visualization>}\}$
\State Initialize text-chart interleaved report $CT_0 \gets \emptyset$; $i \gets 0$
\State Get data overview $DO$ via Equation~\ref{eq: overview}.
\State Update memory of $A$ to $A'$ with  $DO$ and $D$
\State Get report outline $O$ via Equation~\ref{eqa: outline}
\State Finished $\gets$ \textbf{false}
\State Generation step $i=0$
\While{finished is \textbf{false}}
    \State Get step content $x_i={t_i,q_i}$ via Equation~\ref{eqa: writer_output}.
    \State $CT_i \gets CT_{i-1} \cup \{t_i\}$
    \If{$\mathtt{</visualization>}$ in $x_i$} 
        \State Get $q_i$ from $x_i$
        \State Get chart $vc_i$ and $s_i$  as visualization evidence $v_i$ via Equation~\ref{eqa: analysis_call}
        \State $CT_i \gets CT_i \cup \{v_i\}$
       
    \ElsIf{\texttt{<EOS>} in $t_i$}
        \State finished $\gets$ \textbf{true}
    \EndIf
    \State $i \gets i + 1$
\EndWhile
\State \Return $CT$
\end{algorithmic}
\end{algorithm}

\section{Experiment}

\subsection{Experiment Setting}

\textbf{Dataset.} We curate a benchmark of \emph{table-analysis request} pairs from three mainstream report sources to evaluate \methodname{}. Specifically, we sample 20 reports from each source, including Tableau Public Stories~\cite{Tableau}, Our World in Data~\cite{OWID}, and USAFacts~\cite{usafact}, covering 18 topics in total. For each report, we extract the underlying data tables and the original report title as the analysis request. More detailed statistics about the collected instances are provided in Appendix~\ref{sup: datasets}.

\begin{table}[t]
\centering
\caption{Automatic Ranking Results under Qwen3-235B-A22B-Instruct-2507 and Qwen2.5-VL-72B-Instruct.}
\renewcommand{\arraystretch}{1.1}
\fontsize{8}{9}\selectfont 
\setlength{\tabcolsep}{4.4pt}
\begin{tabular}{lllcccc}
\hline

Dataset & Level & Metrics & Direct & DN. & DA. & \cellcolor{pink!30}Ours \\
\hline
\multirow{6}{*}{Tableau}
 & \multirow{2}{*}{chart}    & Read.     & 3.10 & 1.85 & 3.60 & \cellcolor{pink!30}\textbf{1.45} \\
 &                            & Layout.   & 2.95 & \textbf{1.35} & 3.90 & \cellcolor{pink!30}1.80 \\
 & \multirow{2}{*}{chapter}  & T-C Cons. & 1.95 & 2.35 & 4.00 & \cellcolor{pink!30}\textbf{1.70} \\
 &                            & Depth.    & 2.15 & 2.30 & 4.00 & \cellcolor{pink!30}\textbf{1.55} \\
 & \multirow{2}{*}{report}   & Info.     & 2.90 & 1.75 & 3.75 & \cellcolor{pink!30}\textbf{1.60} \\
 &                            & Vis Cons. & 2.55 & 2.60 & 3.05 & \cellcolor{pink!30}\textbf{1.80} \\
\hline
\multirow{6}{*}{\shortstack[l]{OurWorld\\InData}}
 & \multirow{2}{*}{chart}    & Read.     & 2.85 & 2.12 & 3.70 & \cellcolor{pink!30}\textbf{1.30} \\
 &                            & Layout.   & 2.85 & 1.75 & 3.90 & \cellcolor{pink!30}\textbf{1.50} \\
 & \multirow{2}{*}{chapter}  & T-C Cons. & 1.95 & 2.35 & 4.00 & \cellcolor{pink!30}\textbf{1.70} \\
 &                            & Depth.    & 2.20 & 2.40 & 4.00 & \cellcolor{pink!30}\textbf{1.40} \\
 & \multirow{2}{*}{report}   & Info.     & 2.75 & 2.25 & 3.50 & \cellcolor{pink!30}\textbf{1.50} \\
 &                            & Vis Cons. & 2.05 & 2.85 & 3.45 & \cellcolor{pink!30}\textbf{1.65} \\
\hline
\multirow{6}{*}{USAFact}
 & \multirow{2}{*}{chart}    & Read.     & 2.85 & 1.90 & 3.85 & \cellcolor{pink!30}\textbf{1.40} \\
 &                            & Layout.   & 2.80 & 1.65 & 3.95 & \cellcolor{pink!30}\textbf{1.60} \\
 & \multirow{2}{*}{chapter}  & T-C Cons. & 2.10 & 2.05 & 4.00 & \cellcolor{pink!30}\textbf{1.85} \\
 &                            & Depth.    & 2.80 & 2.10 & 4.00 & \cellcolor{pink!30}\textbf{1.10} \\
 & \multirow{2}{*}{report}   & Info.     & 2.90 & 2.25 & 3.45 & \cellcolor{pink!30}\textbf{1.40} \\
 &                            & Vis Cons. & 2.25 & 2.65 & 3.20 & \cellcolor{pink!30}\textbf{1.70} \\
\hline
\end{tabular}
\label{tab:model_eval_setting1_single_col}
\end{table}

\textbf{Baselines.}
We compare with methods from different generation paradigms for text-chart interleaved report with generation by only LLM itself. For direct generation, we prompt LLM to produce the full report in a single pass with \texttt{<visualization>} specifications that are rendered into charts. Furthermore, we include DataNarrative~\cite{islam-etal-2024-datanarrative}, a representative method for the text-first-graph-second paradigm and DeepAnalyze~\cite{deepanalyze}, a representative method for the graph-first-text-second paradigm. For brevity, we denote direct generation by LLMs as \textbf{Direct.}, DataNarrative as \textbf{DN.} and DeepAnalyze as \textbf{DA.}. More details are in Appendix~\ref{sup: baslines}.

\textbf{Metrics.} Given the multimodal nature of generated reports, we evaluate quality at three levels: (i) chart, (ii) chapter pair, and (iii) report.For each level, we adopt two criteria and more details are in Appendix~\ref{sup: metrixs}. The overview is as follows: 

\noindent\textbf{(i) Chart level.}

\textbf{Layout Rationality (Layout.)}: Whether the chart layout uses space effectively.

\textbf{Readability (Read.)}: Whether visual elements are clear and unambiguous to read.

\noindent\textbf{(ii) Chapter level.}

\textbf{Text-Chart Consistency (T-C Cons.)}: Whether the paired text is supported by the chart.

\textbf{Textual Information Depth (Depth.)}: Whether the text goes beyond surface description to provide meaningful, intent-aligned analytical insights.

\noindent\textbf{(iii) Report level.}

\textbf{Informativeness (Info.)}: Whether the report sufficiently addresses the requested analysis with comprehensive and useful content.

\textbf{Visual Consistency (Vis Cons.)}: Whether all charts in the report are consistent in visual style.

\textbf{Evaluation Strategy.}
We employ both an LLM-as-judge (base on GPT-4.1~\cite{openai2024gpt4ocard}) and human evaluators to compare methods. Rather than absolute scoring, we adopt a \textbf{ranking-based} strategy to reduce calibration and stability issues~\cite{dome}. For each instance, evaluators rank the outputs by relative preference and we report the \textbf{average rank result} as the measurement of performance. More details are in Appendix~\ref{sup: eval_details}.

\textbf{Implementations of \methodname.} \methodname{} is implemented as an agentic framework in which the Data-Augmented Analysis Agent and the write-time evidence-constructing writer are based on a multimodal LLM by independent memory. Partial tasks such as outline generation are replaced by LLM with the same contextual memory. The visualization tool $T$ is a program where LLM or MLLM serves as the code generator (details are in Appendix~\ref{sup:vis_tool}). Following the setting by~\citeauthor{yang2025multimodaldeepresearchergeneratingtextchart}, our experiments encompass two model configurations: (1) small scale: Qwen3-VL-32B-Instruct~\cite{Qwen3-VL} serving as the only base model; (2) large hybrid scale:  Qwen3-235B-A22B-Instruct-2507~\cite{qwen3} for text generation and Qwen2.5-VL-72B-Instruct~\cite{Qwen2.5-VL} for multimodal understanding. For fair comparison, all the settings are consistent in both~\methodname{} and baselines. More details are in Appendix~\ref{sup: baslines}.

\subsection{Automatic Report Evaluation}

\begin{table}[t]
\centering
\caption{Automatic Ranking Results under Qwen3-VL-32B-Instruct.}
\renewcommand{\arraystretch}{1.1}
\fontsize{8}{9}\selectfont
\setlength{\tabcolsep}{4.4pt}
\begin{tabular}{lllcccc}
\hline
Dataset & Level & Metrics & Direct & DN. & DA. & \cellcolor{pink!30}Ours \\
\hline
\multirow{6}{*}{Tableau}
 & \multirow{2}{*}{chart}    & Read.     & 3.05 & \textbf{1.45} & 3.65 & \cellcolor{pink!30}1.85 \\
 &                            & Layout.   & 2.80 & \textbf{1.45} & 3.85 & \cellcolor{pink!30}1.90 \\
 & \multirow{2}{*}{chapter}  & T-C Cons. & 2.05 & 2.70 & 4.00 & \cellcolor{pink!30}\textbf{1.25} \\
 &                            & Depth.    & 2.35 & 2.45 & 4.00 & \cellcolor{pink!30}\textbf{1.20} \\
 & \multirow{2}{*}{report}   & Info.     & 3.05 & 1.95 & 3.55 & \cellcolor{pink!30}\textbf{1.45} \\
 &                            & Vis Cons. & 2.70 & 2.50 & 3.10 & \cellcolor{pink!30}\textbf{1.70} \\
\hline
\multirow{6}{*}{\shortstack[l]{OurWorld\\InData}}
 & \multirow{2}{*}{chart}    & Read.     & 3.30 & 1.80 & 3.40 & \cellcolor{pink!30}\textbf{1.50} \\
 &                            & Layout.   & 3.25 & \textbf{1.40} & 3.60 & \cellcolor{pink!30}1.70 \\
 & \multirow{2}{*}{chapter}  & T-C Cons. & 2.20 & 2.75 & 4.00 & \cellcolor{pink!30}\textbf{1.05} \\
 &                            & Depth.    & 2.55 & 2.20 & 4.00 & \cellcolor{pink!30}\textbf{1.25} \\
 & \multirow{2}{*}{report}   & Info.     & 3.10 & 2.20 & 3.35 & \cellcolor{pink!30}\textbf{1.35} \\
 &                            & Vis Cons. & 2.55 & 2.85 & 2.70 & \cellcolor{pink!30}\textbf{1.50} \\
\hline
\multirow{6}{*}{USAFact}
 & \multirow{2}{*}{chart}    & Read.     & 3.30 & 1.80 & 3.40 & \cellcolor{pink!30}\textbf{1.50} \\
 &                            & Layout.   & 2.95 & \textbf{1.30} & 3.65 & \cellcolor{pink!30}2.10 \\
 & \multirow{2}{*}{chapter}  & T-C Cons. & 2.65 & 2.20 & 4.00 & \cellcolor{pink!30}\textbf{1.15} \\
 &                            & Depth.    & 2.50 & 2.40 & 4.00 & \cellcolor{pink!30}\textbf{1.10} \\
 & \multirow{2}{*}{report}   & Info.     & 3.50 & 1.75 & 3.25 & \cellcolor{pink!30}\textbf{1.50} \\
 &                            & Vis Cons. & 2.50 & 2.50 & 3.15 & \cellcolor{pink!30}\textbf{1.65} \\
\hline
\end{tabular}
\label{tab:model_eval_setting2}
\end{table}

\textbf{\methodname{} demonstrates consistent, multi-level advantages from chart quality to overall report quality.} We report the ranking result on all metrics by GPT-4.1 in Table~\ref{tab:model_eval_setting1_single_col}.~\methodname{} outperforms all baselines across all datasets from all evaluation levels. The staged baselines \textbf{DN.} and \textbf{DA.} receive worse rankings on the chart-text consistency (\textbf{T-C Cons.}) and information depth (\textbf{Depth.}), indicating that their generated narratives are often weakly grounded in the final visual evidence and tend to remain at a surface descriptive level.
We attribute the advantages to the following factors: (1) decoupling the non-trivial visualization/analysis workload into two agents, avoiding redundant prompting and weak grounding during generation; (2) an outline-guided, writing-time interleaving process that issues on-demand visualization requests and injects returned charts back into the generated context, ensuring that subsequent claims are directly constrained by concrete visual evidence.

\textbf{~\methodname{} remains consistently effective across base models, indicating cross model-configuration scalability.} As illustrated in Table~\ref{tab:model_eval_setting1_single_col} and Table~\ref{tab:model_eval_setting2}, \methodname{} consistently outperforms all baselines under both model configurations and across multi-level metrics, suggesting that its gains do not rely on a particular backbone choice.
We attribute this cross-model robustness to our \emph{training-free} framework design and the use of model-agnostic prompts in each module, which makes the pipeline less sensitive to backbone capacity while preserving the same decomposition and grounding behaviors across models.

\subsection{Human Evaluation}

\begin{table}[t]
\centering
\caption{Human Ranking Results when applying Qwen3-235B-A22B-Instruct-2507 \& Qwen2.5-VL-72B-Instruct as base models.}
\label{tab: human_eval1}
\renewcommand{\arraystretch}{1.1}
\fontsize{8}{9}\selectfont
\setlength{\tabcolsep}{4.4pt}
\begin{tabular}{lllcccc}
\hline
Dataset & Level & Metrics & Direct & DN. & DA. & \cellcolor{pink!30}Ours \\
\hline
\multirow{6}{*}{Tableau}
 & \multirow{2}{*}{chart}      & Read.     & 3.10 & 2.27 & 3.40 & \cellcolor{pink!30}\textbf{1.23} \\
 &                             & Layout.   & 2.93 & 2.23 & 3.53 & \cellcolor{pink!30}\textbf{1.30} \\
 & \multirow{2}{*}{chapter}    & T-C Cons. & 2.23 & 2.70 & 4.00 & \cellcolor{pink!30}\textbf{1.07} \\
 &                             & Depth.    & 3.20 & 2.30 & 3.37 & \cellcolor{pink!30}\textbf{1.13} \\
 & \multirow{2}{*}{report}     & Info.     & 3.60 & 2.10 & 3.07 & \cellcolor{pink!30}\textbf{1.23} \\
 &                             & Vis Cons. & 2.70 & 2.70 & 3.20 & \cellcolor{pink!30}\textbf{1.40} \\
\hline
\multirow{6}{*}{\shortstack[l]{OurWorld\\InData}}
 & \multirow{2}{*}{chart}      & Read.     & 3.23 & 2.27 & 3.17 & \cellcolor{pink!30}\textbf{1.33} \\
 &                             & Layout.   & 3.07 & 2.20 & 3.50 & \cellcolor{pink!30}\textbf{1.23} \\
 & \multirow{2}{*}{chapter}    & T-C Cons. & 2.40 & 2.60 & 4.00 & \cellcolor{pink!30}\textbf{1.00} \\
 &                             & Depth.    & 2.80 & 2.60 & 3.57 & \cellcolor{pink!30}\textbf{1.03} \\
 & \multirow{2}{*}{report}     & Info.     & 3.33 & 2.23 & 3.20 & \cellcolor{pink!30}\textbf{1.23} \\
 &                             & Vis Cons. & 2.77 & 2.53 & 3.23 & \cellcolor{pink!30}\textbf{1.47} \\
\hline
\multirow{6}{*}{USAFacts}
 & \multirow{2}{*}{chart}      & Read.     & 3.20 & 2.17 & 3.47 & \cellcolor{pink!30}\textbf{1.17} \\
 &                             & Layout.   & 2.83 & 2.23 & 3.67 & \cellcolor{pink!30}\textbf{1.27} \\
 & \multirow{2}{*}{chapter}    & T-C Cons. & 2.37 & 2.70 & 3.87 & \cellcolor{pink!30}\textbf{1.07} \\
 &                             & Depth.    & 2.90 & 2.33 & 3.70 & \cellcolor{pink!30}\textbf{1.07} \\
 & \multirow{2}{*}{report}     & Info.     & 3.40 & 2.07 & 3.33 & \cellcolor{pink!30}\textbf{1.20} \\
 &                             & Vis Cons. & 2.57 & 2.80 & 3.43 & \cellcolor{pink!30}\textbf{1.20} \\
\hline
\end{tabular}
\end{table}

\textbf{Human preferences are consistent with LLM-as-a-judge rankings.} Comparing Table~\ref{tab:model_eval_setting1_single_col} with Table~\ref{tab: human_eval1} and Table~\ref{tab:model_eval_setting2} with Table~\ref{tab: human_eval2}, the ranking result under the same mode setting by LLM-as-a-Judge and human evaluators, \methodname{} is ranked best under both model configurations in most cases, while DataNarrative and DeepAnalyze follow behind with the only exception at the report level result on Tableau. 
Under the large hybrid setting (Table~\ref{tab: human_eval1}), \methodname{} achieves near-top ranks across all three sources and six criteria, including strong chapter-level text-chart consistency and depth and report-level informativeness and visual consistency. Similar pattern holds in the smaller model setting (Table~\ref{tab: human_eval2}), where \methodname{} again ranks best on most metrics. The only notable deviation is the \emph{report-level} results on Tableau, where DataNarrative outranks \methodname{} on informativeness and visual consistency.  Collectively, the high qualitative agreement supports the reasonableness of LLM-as-a-Judge as a scalable proxy for human preference in multi-level report evaluation, and it concurrently strengthens the credibility of \methodname{}'s reported multi-level advantages beyond a single evaluator type. 

\begin{table}[!t]
\centering
\caption{Human Ranking Results when applying Qwen3-VL-32B-Instruct as base model.}
\label{tab: human_eval2}
\renewcommand{\arraystretch}{1.1}
\fontsize{8}{9}\selectfont
\setlength{\tabcolsep}{4.4pt}
\begin{tabular}{lllcccc}
\hline
Dataset & Level & Metrics & Direct & DN. & DA. & \cellcolor{pink!30}Ours \\
\hline
\multirow{6}{*}{Tableau}
 & \multirow{2}{*}{chart}      & Read.     & 3.60 & 2.00 & 2.50 & \cellcolor{pink!30}\textbf{1.90} \\
 &                             & Layout.   & 3.80 & 1.90 & 2.60 & \cellcolor{pink!30}\textbf{1.70} \\
 & \multirow{2}{*}{chapter}    & T-C Cons. & 3.70 & 2.10 & 2.40 & \cellcolor{pink!30}\textbf{1.80} \\
 &                             & Depth.    & 3.90 & 2.30 & 2.30 & \cellcolor{pink!30}\textbf{1.50} \\
 & \multirow{2}{*}{report}     & Info.     & 3.90 & \textbf{1.60} & 2.40 & \cellcolor{pink!30}2.10 \\
 &                             & Vis Cons. & 3.60 & \textbf{1.40} & 2.60 & \cellcolor{pink!30}2.40 \\
\hline
\multirow{6}{*}{\shortstack[l]{OurWorld\\InData}}
 & \multirow{2}{*}{chart}      & Read.     & 3.70 & 2.70 & 2.10 & \cellcolor{pink!30}\textbf{1.50} \\
 &                             & Layout.   & 3.90 & 2.30 & 2.20 & \cellcolor{pink!30}\textbf{1.60} \\
 & \multirow{2}{*}{chapter}    & T-C Cons. & 3.00 & 2.80 & 2.90 & \cellcolor{pink!30}\textbf{1.30} \\
 &                             & Depth.    & 3.90 & 2.30 & 2.10 & \cellcolor{pink!30}\textbf{1.70} \\
 & \multirow{2}{*}{report}     & Info.     & 3.80 & 2.30 & 2.00 & \cellcolor{pink!30}\textbf{1.90} \\
 &                             & Vis Cons. & 3.40 & 2.30 & 2.30 & \cellcolor{pink!30}\textbf{2.00} \\
\hline
\multirow{6}{*}{USAFacts}
 & \multirow{2}{*}{chart}      & Read.     & 3.50 & 2.70 & 2.30 & \cellcolor{pink!30}\textbf{1.80} \\
 &                             & Layout.   & 3.70 & 2.20 & \textbf{2.00} & \cellcolor{pink!30}2.30 \\
 & \multirow{2}{*}{chapter}    & T-C Cons. & 3.60 & 2.40 & \textbf{1.80} & \cellcolor{pink!30}\textbf{1.80} \\
 &                             & Depth.    & 3.90 & 2.50 & 2.20 & \cellcolor{pink!30}\textbf{1.80} \\
 & \multirow{2}{*}{report}     & Info.     & 3.90 & 1.90 & 2.10 & \cellcolor{pink!30}\textbf{1.50} \\
 &                             & Vis Cons. & 3.60 & 1.90 & 3.00 & \cellcolor{pink!30}\textbf{1.50} \\
\hline
\end{tabular}
\end{table}

\textbf{\methodname{} produces reports with higher practical utility under human evaluation.}
From Table~\ref{tab: human_eval1} and Table~\ref{tab: human_eval2}, \methodname{} is ranked best in most cases evaluated by different human evaluators. On OurWorldInData, \methodname{} ranks 1.30 in text-chart consistency and 1.90 in information richness, whereas directly using base mode ranks 3.00 in text-chart consistency and 3.80 in information richness. These advantages support~\methodname{} produces more informative, data-driven reports with high-quality visualization for human readers in practice.

\subsection{Ablation Study}
\begin{table}[!t]
\centering
\caption{The Ablation Study of \methodname{} with Qwen3-VL-32B-Instruct.}
\label{tab: abs_study}
\renewcommand{\arraystretch}{1.1}
\fontsize{8}{9}\selectfont
\setlength{\tabcolsep}{2pt}
\begin{tabular}{lllcccc}
\hline
Dataset & Level & Metrics & Direct & w/o Vis. & w/o Writer. & \cellcolor{pink!30}Ours \\
\hline
\multirow{6}{*}{Tableau}
 & \multirow{2}{*}{chart}      & Read.     & 3.60 & 2.00 & 2.50 & \cellcolor{pink!30}\textbf{1.90} \\
 &                             & Layout.   & 3.80 & 1.90 & 2.60 & \cellcolor{pink!30}\textbf{1.70} \\
 & \multirow{2}{*}{chapter}    & T-C Cons. & 3.70 & 2.10 & 2.40 & \cellcolor{pink!30}\textbf{1.80} \\
 &                             & Depth.    & 3.90 & 2.30 & 2.30 & \cellcolor{pink!30}\textbf{1.50} \\
 & \multirow{2}{*}{report}     & Info.     & 3.90 & 2.10 & 2.40 & \cellcolor{pink!30}\textbf{1.60} \\
 &                             & Vis Cons. & 3.60 & 2.40 & 2.60 & \cellcolor{pink!30}\textbf{1.40} \\
\hline
\multirow{6}{*}{\shortstack[l]{OurWorld\\InData}}
 & \multirow{2}{*}{chart}      & Read.     & 3.70 & 2.70 & 2.10 & \cellcolor{pink!30}\textbf{1.50} \\
 &                             & Layout.   & 3.90 & 2.30 & 2.20 & \cellcolor{pink!30}\textbf{1.60} \\
 & \multirow{2}{*}{chapter}    & T-C Cons. & 3.00 & 2.80 & 2.90 & \cellcolor{pink!30}\textbf{1.30} \\
 &                             & Depth.    & 3.90 & 2.30 & 2.10 & \cellcolor{pink!30}\textbf{1.70} \\
 & \multirow{2}{*}{report}     & Info.     & 3.80 & 2.30 & 2.00 & \cellcolor{pink!30}\textbf{1.90} \\
 &                             & Vis Cons. & 3.40 & 2.30 & 2.30 & \cellcolor{pink!30}\textbf{2.00} \\
\hline
\multirow{6}{*}{USAFacts}
 & \multirow{2}{*}{chart}      & Read.     & 3.50 & 2.70 & 2.00 & \cellcolor{pink!30}\textbf{1.80} \\
 &                             & Layout.   & 3.70 & 2.20 & 2.30 & \cellcolor{pink!30}\textbf{1.80} \\
 & \multirow{2}{*}{chapter}    & T-C Cons. & 3.60 & 2.40 & 2.20 & \cellcolor{pink!30}\textbf{1.80} \\
 &                             & Depth.    & 3.90 & 2.50 & 2.10 & \cellcolor{pink!30}\textbf{1.50} \\
 & \multirow{2}{*}{report}     & Info.     & 3.90 & 1.90 & 2.60 & \cellcolor{pink!30}\textbf{1.60} \\
 &                             & Vis Cons. & 3.60 & 1.90 & 3.00 & \cellcolor{pink!30}\textbf{1.50} \\
\hline
\end{tabular}
\end{table}

\textbf{Effectiveness of the Data-Augmented Analysis Agent.}
We ablate the Data-Augmented Analysis Agent by removing $A'$ and forcing the writer to couple visualization planning and execution with long-form generation under the same inputs ($R_{\text{user}}$ and $DO$).
As shown in Table~\ref{tab: abs_study}, this variant degrades chart-level quality and chart-grounded reasoning, with clear drops on \textbf{Read.} and \textbf{Depth}.
For example, on USAFacts, \textbf{Read.} worsens from 1.8 to 2.7 and \textbf{Depth} deteriorates from 1.5 to 2.5 after removing $A'$.
These results highlight the benefit of decoupling non-trivial visualization decisions from narrative generation. A dedicated analysis module is critical for reliably selecting variables/aggregations and producing grounded visual evidence, freeing the writer to focus on coherent narration and enabling deeper, chart-supported insights.

\textbf{Effectiveness of real-time data interaction during writing.}
We ablate write-time evidence construction by disabling real-time interaction, reducing \methodname{} to a \emph{text-first--graph-second} pipeline that the writer first generates the full narrative with deferred visualization placeholders, and the analysis agent is invoked only after the text is finalized to produce the corresponding charts. As shown in Table~\ref{tab: abs_study}, this ablation primarily harms chart-text alignment and textual analysis depth over report. On Tableau the rank of T-C Cons. worsens from 1.8 to 2.4 and Textual Depth degrades from 1.5 to 2.3 after removing the interaction mechanism. This suggests that visual evidence construction at write time through data interaction during writing process is essential for continuously grounding claims on newly generated visuals, allowing the writer to iteratively request targeted evidence, correct mismatches, and produce informative content.

\subsection{More Discussion}

\begin{table}[t]
\centering
\setlength{\tabcolsep}{4pt} 
\caption{Report-level statistics when applying Qwen3-235B-A22B-Instruct-2507 \& Qwen2.5-VL-72B-Instruct as base models. Fig. denotes the average number of figures per report, Words. denotes the average tokens per report and Content. measures the amount of information in the report text.}
\renewcommand{\arraystretch}{1.1}
\fontsize{8}{9}\selectfont 
\setlength{\tabcolsep}{2.5pt}
\begin{tabular}{lccccc}  
\toprule 
Dataset & Metrics & Direct & DN. & DA. & \cellcolor{pink!30}Ours \\
\midrule 

\multirow{3}{*}{Tableau} &Fig. & 3.85 & 6.10 & 4.05 & \cellcolor{pink!30}\textbf{6.90} \\
&Words. & 879.60 & 1203.00 & 1852.80 & \cellcolor{pink!30}\textbf{2004.45} \\
&Content. & 16.35 & 12.60 & 15.60 & \cellcolor{pink!30}\textbf{18.85} \\

\multirow{3}{*}{OurWorldInData} &Fig. & 3.70 & 7.20 & 4.15 & \cellcolor{pink!30}\textbf{8.20} \\
&Words.& 1037.10 & 1667.65 & 1779.90 & \cellcolor{pink!30}\textbf{2132.60} \\
&Content.& 14.95 & 16.70 & 15.35 & \cellcolor{pink!30}\textbf{18.40} \\

\multirow{3}{*}{USAFact} &Fig. & 3.55 & 5.95 & 3.10 & \cellcolor{pink!30}\textbf{6.55} \\
&Words.& 924.30 & 1695.90 & 1755.10 & \cellcolor{pink!30}\textbf{2310.10} \\
&Content. & 15.65 & 15.00 & 15.90 & \cellcolor{pink!30}\textbf{20.40} \\
\bottomrule
\end{tabular}
\label{tab: more_discussion1}
\end{table}

\begin{table}[t]
\centering
\setlength{\tabcolsep}{4pt} 
\caption{Report-level statistics when applying Qwen3-VL-32B-Instruct as base model. Other details are the same of Table~\ref{tab: more_discussion1}.}
\renewcommand{\arraystretch}{1.1}
\fontsize{8}{9}\selectfont 
\setlength{\tabcolsep}{2.5pt}
\begin{tabular}{lccccc}  
\toprule 
Dataset & Metrics & Direct & DN. & DA. & \cellcolor{pink!30}Ours \\
\midrule 

\multirow{3}{*}{Tableau} &Fig. & 3.00 & 7.60 & 4.05 & \cellcolor{pink!30}\textbf{8.30} \\
&Words. & 1154.25 & 1694.95 & 2004.45 & \cellcolor{pink!30}\textbf{4071.05} \\
&Content. & 11.85 & 16.45 & 15.60 & \cellcolor{pink!30}\textbf{18.85} \\

\multirow{3}{*}{OurWorldInData} &Fig. & 3.05 & 6.55 & 4.15 & \cellcolor{pink!30}\textbf{8.80} \\
&Words.& 1539.10 & 2542.20 & 1779.90 & \cellcolor{pink!30}\textbf{4300.15} \\
&Content.& 14.40 & 13.80 & 15.35 & \cellcolor{pink!30}\textbf{18.00} \\

\multirow{3}{*}{USAFact} &Fig. & 2.55 & 5.90 & 3.10 & \cellcolor{pink!30}\textbf{6.55} \\
&Words.& 1117.35 & 1593.60 & 1755.10 & \cellcolor{pink!30}\textbf{2608.10} \\
&Content. & 14.85 & 16.45 & 15.90 & \cellcolor{pink!30}\textbf{16.85} \\
\bottomrule
\end{tabular}
\label{tab: more_discussion2}
\end{table}

\textbf{\methodname{} produces visualization-rich and long reports.}
From Table~\ref{tab: more_discussion1} and Table~\ref{tab: more_discussion2}, \methodname{} generates the most visualization and the longest reports across all datasets. on Tableau,the reports generated by~\methodname{} contain 8.30 visualization with 4071.05 tokens averagely while the reports generated by DataNarrative contain 7.60 visualizations with 1694.95 tokens. These results indicate that real-time data interaction sustains evidence acquisition during writing support both comprehensive generation of visualization and contextual content.

\textbf{\methodname{} contains the most information in reports.} From Table~\ref{tab: more_discussion1} and Table~\ref{tab: more_discussion2}, beyond visualization and length, the reports from~\methodname{} contain more information than all baselines across datasets. With the same information extraction setting (see Appendix~\ref{sup: content_extract} for extraction details),~\methodname{} attains highest Content. score. On OurWorldInData, ~\methodname{} contains 18 pieces of information while Direct contains 14.40 pieces. This suggests that the additional figures and tokens translate into denser, more informative, and decision-oriented analysis.

\section{Conclusion}
This paper reframes data-driven report generation as a writing-time evidence construction problem: staged text–chart pipelines (text-first or graph-first) tend to "freeze" intermediate insights and charts, making it difficult to maintain chart–text consistency as the narrative develops. We propose \methodname{}, a training-free text–chart interleaving framework that couples a data-augmented analysis agent with a real-time interactive writer, allowing the model to request and inject grounded visual evidence exactly when needed during drafting. Experiments across multiple multi-table benchmarks show that this interleaving paradigm produces reports with stronger chart quality, tighter chart–text alignment, and higher overall usefulness, suggesting a general paradigm for grounded long-form generation where evidence is treated as a first-class, dynamically constructed context.

\bibliography{custom}

\newpage
\appendix

\section{Limitation}

\label{sec:limitation}

\begin{table}[h]
\centering
\caption{Cost comparison across methods using Qwen3-VL-32B-Instruct. API. denotes the average number of API calls per report, Latency. (s) denotes average generation time in seconds.}
\small
\setlength{\tabcolsep}{4pt}
\resizebox{0.48\textwidth}{!}{
\begin{tabular}{lcccc}
\toprule
\multirow{2}{*}{Dataset} & \multicolumn{2}{c}{Direct} & \multicolumn{2}{c}{DN.} \\
\cmidrule(lr){2-3} \cmidrule(lr){4-5}
& API. & Latency. (s) & API. & Latency. (s) \\
\midrule
Tableau & 3.95 & 145.50 & 14.30 & 792.19 \\
OurWorldInData & 3.40 & 179.33 & 12.60 & 1047.89 \\
USAFact & 3.65 & 130.33 & 11.90 & 725.44 \\
\midrule
\multirow{2}{*}{Dataset} & \multicolumn{2}{c}{DA.} & \multicolumn{2}{c}{\cellcolor{pink!30}Ours} \\
\cmidrule(lr){2-3} \cmidrule(lr){4-5}
& API. & Latency. (s) & \cellcolor{pink!30}API. & \cellcolor{pink!30}Latency. (s) \\
\midrule
Tableau & 18.93 & 79.50 & \cellcolor{pink!30}41.85 & \cellcolor{pink!30}946.15 \\
OurWorldInData & 15.75 & 78.05 & \cellcolor{pink!30}43.20 & \cellcolor{pink!30}1511.40 \\
USAFact & 21.32 & 81.80 & \cellcolor{pink!30}33.13 & \cellcolor{pink!30}1034.60 \\
\bottomrule
\end{tabular}
}
\label{tab: cost_analysis}
\end{table}

\textbf{Additional time and API cost for write-time visual evidence construction.}
Write-time evidence construction increases tool usage and end-to-end latency because the writer may trigger multiple visualization requests during generation.
As shown in Table~\ref{tab: cost_analysis}, \methodname{} incurs extra API calls mainly from (i) constructing the dataset overview $DO$ and (ii) servicing each visualization request, which includes analyzing intent and specification, tool-based rendering (with iterative refinement) and grounded caption generation.
Despite this overhead, \methodname{} achieves substantially better report quality ( in Tables~\ref{tab:model_eval_setting1_single_col} and~\ref{tab:model_eval_setting2}).
In practice, engineering optimizations such as caching or reusing intermediate aggregations, batching compatible requests, and parallel tool execution further reduce the cost while preserving evidence fidelity.

\textbf{Robust visualization execution and recovery.}
Since~\methodname{} needs code execution to render visualization result, repeated execution failures may reduce the amount and quality of visual evidence in long reports (see the case study in Appendix~\ref{sup: case study}).
Future work can improve robustness via verified execution and self-repair, such as type-safe data adapters, unit checks on aggregations, constrained chart templates, as well as graceful fallback strategies that still return minimally informative visuals under partial failures.

\section{The Details of Experiment Setting}
\label{sup: experiment setting}

\subsection{The details of datasets.}
\label{sup: datasets}

We construct a comprehensive evaluation dataset by curating reports from three authoritative data sources renowned for their reliability and analytical depth. We collect 20 reports from each of the three sources: Tableau's public data gallery~\cite{Tableau}, Our World in Data~\cite{OWID}, and USAFact~\cite{usafact}, resulting in a total of 60 reports for evaluation. For each source, we select reports covering diverse thematic domains including environment, economy, health, education, and social issues, ensuring a broad coverage of analytical perspectives. Table~\ref{tab:dataset_topics} presents the detailed topic distribution across the three data sources, where each number indicates the count of reports collected for that topic. For each report, we extract the title, the full PDF content, and the underlying tabular data, which is then standardized and stored in CSV format to facilitate consistent processing across all methods.

To provide a comprehensive understanding of the data scale and complexity, Table~\ref{tab:dataset_stats} presents detailed statistics for each dataset. The ``Table.'' column indicates the total number of tables contained across all reports within each dataset. The ``Row.'' column reports the average total number of rows across all tables per report, reflecting the data volume and granularity. The ``Column.'' column shows the average total number of columns across all tables per report, representing the dimensionality and feature richness of the data.

\begin{table}[t]
\centering
\small
\caption{Topic distribution across the three data sources.}
\label{tab:dataset_topics}
\fontsize{8}{9}\selectfont
\setlength{\tabcolsep}{4pt}
\begin{tabular}{lccc}
\toprule
\textbf{Topic} & \textbf{Tableau} & \textbf{OurWorldInData} & \textbf{USAFact} \\
\midrule
Environment & 3 & 2 & 2 \\
Economy & 3 & 2 & 2 \\
Health & 1 & 2 & 3 \\
Business & 1 & - & - \\
Social & 2 & 2 & 1 \\
Transport & 2 & - & - \\
Energy & 1 & - & - \\
Education & 4 & 2 & 3 \\
Religion & 1 & - & - \\
Psychology & 1 & - & - \\
Entertainment & 1 & - & - \\
Population & - & 2 & 2 \\
Food and Agriculture & - & 2 & - \\
AI & - & 2 & - \\
Human Rights & - & 2 & - \\
War & - & 2 & 3 \\
Government & - & - & 2 \\
Crime & - & - & 2 \\
\midrule
\textbf{Total} & \textbf{20} & \textbf{20} & \textbf{20} \\
\bottomrule
\end{tabular}
\end{table}

\begin{table}[t]
\centering
\caption{Statistics of different datasets.}
\fontsize{8}{9}\selectfont
\setlength{\tabcolsep}{10pt}
\begin{tabular}{lccc}
\hline
Dataset & Table. & Row. & Colum. \\
\hline
Tableau & 113 & 1034.55 & 61.65 \\
OurWorldInData & 109 & 458.1 & 35.3 \\
USAFact & 87 & 295.8 & 37.1 \\
\hline
\end{tabular}
\label{tab:dataset_stats}
\end{table}

\subsection{The details of Implementation}
\label{sup: baslines}

We evaluate two model configurations to assess the scalability and practicality of \methodname{}. The first setting employs Qwen3-VL-32B-Instruct, which is deployed locally on two NVIDIA A800 GPUs (80GB each) using vLLM for efficient inference. The second setting combines Qwen3-235B-A22B-Instruct-2507 for text generation with Qwen2.5-VL-72B-Instruct for multimodal understanding, both accessed via the OpenRouter API platform~\citep{openrouter}. This hybrid configuration demonstrates the framework's flexibility across different deployment scenarios while maintaining consistent performance.

 For a controlled comparison, we use the same plotting backend for all methods that emit visualization specifications, and we keep decoding settings consistent across all LLM/MLLM calls (temperature set to 0).
\begin{itemize}
    \item Direct Prompting: This baseline serves as a straightforward, end-to-end approach where we directly concatenate the user's intent with the full tabular data (serialized into a text format) as the input prompt for the Large Language Model (LLM). Without relying on intermediate reasoning steps or specialized modules, the LLM is instructed to generate a comprehensive report that embeds <visualization> tags at appropriate locations. Subsequently, a deterministic post-processing module parses these tags and employs a rendering method to transform the specified contents into actual chart images.
    
    \item DataNarrative: We adopt the established architecture proposed by \citet{islam-etal-2024-datanarrative} to serve as another baseline. Following the original implementation, this method generates reports conditioned on the input data and narrative goals. Similar to the Direct Prompting, the method generates <visualization> tags in final report text. To ensure a fair comparison and isolate the generation quality from the rendering quality, we utilize the identical plotting tool used in the Direct Prompting baseline to convert these tag contents into visual chart.
    
    \item DeepAnalyze: We implement the graph-first-text-second paradigm following~\citet{deepanalyze}, which introduces the first agentic LLM specifically designed for autonomous data science. DeepAnalyze performs comprehensive data analysis and visualization before report generation by employing specialized action tokens (\texttt{<Analyze>}, \texttt{<Understand>}, \texttt{<Code>}, \texttt{<Execute>}, \texttt{<Answer>}) to interact with data environments. It follows a curriculum-based training paradigm that progressively acquires and integrates multiple capabilities including data question answering, specialized analytical tasks, and open-ended data research. For fair comparison, we use the same datasets and user intents as inputs to generate analysis and visualizations, then employ its model to produce the final markdown report conditioned on these precomputed visualizations.

    \item \methodname{}: We implement the proposed report generation framework with two distinct model configurations to validate its flexibility across different deployment scenarios. In the first setting, we employ Qwen3-VL-32B-Instruct as the unified model throughout the entire pipeline, handling all tasks including data analysis, chart generation, chart refinement, and final report writing. This single-model configuration demonstrates the framework's capability to deliver coherent reports with a moderately-sized multimodal LLM. In the second setting, we adopt a hybrid architecture that allocates tasks based on their modality requirements: the text-based Data-Augmented Analysis Agent ($A'$) utilizes Qwen3-235B-A22B-Instruct-2507 for analytical reasoning and code generation, while vision-intensive tasks including chart refinement and multimodal report generation are delegated to Qwen2.5-VL-72B-Instruct. This hybrid configuration leverages the strengths of larger language models for complex reasoning while employing specialized multimodal models for visual understanding and generation tasks.

\end{itemize}

\subsection{The details of Evaluation}
\label{sup: eval_details}

\textbf{Ranking-based Protocol.} Rather than assigning absolute scores on a fixed scale, our evaluators are provided with generated reports and detailed rubrics to \textbf{rank} outputs from different methods. This design addresses common limitations of scalar scoring for open-ended generation, including:
\begin{enumerate}
    \item Poor calibration and inconsistent use of rating scales across instances or judges.
    \item Difficulty in making absolute scores comparable across heterogeneous analysis intents.
    \item Sensitivity to anchoring and central-tendency effects.
\end{enumerate}
In contrast, ranking focuses on relative preference under the same input condition, yielding more stable comparisons. We aggregate per-instance rankings across the evaluation set (via average rank) to obtain the final method-level performance for each metric and granularity.

\textbf{The prompt for each metrics.}
\label{sup: metrixs}
We design six evaluation metrics to assess different aspects of report quality. The evaluation prompts for these metrics are presented as follows: Layout Rationality (Figure~\ref{fig:layout_prompt}), Readability (Figure~\ref{fig:read_prompt}), Text-Chart Consistency (Figure~\ref{fig:tc_prompt}), Textual Informative Depth (Figure~\ref{fig:depth_prompt}), Informativeness (Figure~\ref{fig:info_prompt}), and Visual Consistency (Figure~\ref{fig:vis_cons_prompt}). Each prompt provides detailed rubrics and example-based criteria to ensure consistent and meaningful rankings.

\begin{figure*}[htbp]
\begin{tcolorbox}[colback=gray!5!white,colframe=gray!75!black,title=Prompt of Layout. evaluation]

\#\# Task\\
Your task is to evaluate and compare the chart layouts, determining which one best utilizes spatial arrangement to tell a compelling data-driven story.\\

\#\# Layout Evaluation Criteria\\
Layout refers to how charts, text, and graphical elements are orchestrated to guide the reader's understanding. Based on the provided examples, the ideal layout should function like a data journalism piece, prioritizing:\\

- **Narrative-Integrated Flow**: Prefer layouts where the visual hierarchy mirrors the analytical logic. Look for a structure that moves from "Setting the Scene" (descriptive maps/distributions) to "Deep Dives" (scatter plots/trends) and ends with a "Synthesis" (conclusion).\\
- **Embedded Insight \& Annotation**: Prefer layouts that place insights *inside* the chart boundaries. High scores go to layouts using **direct labeling, arrows pointing to outliers, and on-chart text boxes** (e.g., "Significant drop..." or "Inverse Relationship") rather than relying solely on external captions.\\
- **Synthesized Dashboarding**: Prefer reports that utilize a **multi-panel dashboard** layout (typically at the end) to aggregate key metrics (maps, trends, and stats) into a single high-level view for cross-metric comparison.\\
- **Question-Driven Scaffolding**: Prefer layouts where section headers pose a question (e.g., "Is Access Improving?") and the immediately following chart provides the visual answer. The chart titles should be statement-based summaries of the data.\\
- **Statistical \& Visual Consistency**: Prefer layouts that maintain a rigid grid for complex elements—such as aligning **diverging bar charts** or **scatter plots with LOESS curves**—ensuring that reference lines (medians, averages) and error bars are legible and consistent across different figures.\\

\#\# Input\\
I have uploaded the chart pictures below. They are grouped by Report ID.

\end{tcolorbox}
\caption{Prompt for evaluating report chart layout.}
\label{fig:layout_prompt}
\end{figure*}

\begin{figure*}[htbp]
\begin{tcolorbox}[colback=gray!5!white,colframe=gray!75!black,title=Prompt of Read. evaluation]

\#\# Task\\
Your task is to evaluate the quality and readability of chart images from multiple reports.\\

\#\# Evaluation Criteria\\
You will assess how effectively the charts communicate complex analytical findings. Based on the provided examples, high-quality charts should prioritize **statistical depth**, **narrative context**, and **multidimensional synthesis**. Use the following specific criteria:\\

- **Statistical \& Analytical Rigor**: Prefer charts that go beyond raw data points to include statistical enhancements. Look for features such as **trend lines (e.g., LOESS smoothing)** to show correlations, **error bars/confidence intervals** to show variability, or **logarithmic scales** to handle exponential data.\\
- **Narrative-Driven Annotation**: Prefer charts that integrate the "story" directly into the visual. The chart should use **descriptive titles, callout boxes, and direct labeling** of anomalies or key insights (e.g., "Largest decline in delayed care") rather than requiring the reader to hunt for meaning.\\
- **Multidimensional Synthesis (Dashboarding)**: Prefer visualizations that combine multiple related metrics into a single coherent view (e.g., a dashboard combining maps, trend lines, and bar charts). High scores go to layouts that synthesize **geography, frequency, and magnitude/intensity** in one glance.\\
- **Distributional \& Comparative Clarity**: Prefer charts that reveal the *shape* of the data rather than just averages. Look for **box plots** showing spreads, **diverging bar charts** showing positive/negative splits, or **histograms** with reference lines (e.g., "Citywide Median") that allow for immediate benchmarking.\\
- **Geospatial \& Temporal Context**: When location or time is relevant, prefer charts that effectively map data to **geographic clusters** (e.g., bubble maps, choropleth maps) or show clear **temporal evolution** (e.g., distinct pre/post periods or long-term trends) without visual clutter.\\

\#\# Input\\
I have uploaded the chart pictures below. They are grouped by Report ID.

\end{tcolorbox}
\caption{Prompt for evaluating chart readability.}
\label{fig:read_prompt}
\end{figure*}

\begin{figure*}[h]
\begin{tcolorbox}[colback=gray!5!white,colframe=gray!75!black,title=Prompt of T-C Cons. evaluation]

\#\# Task\\
Your task is to evaluate and rank the quality of text-image pairs across multiple reports based on **Text-Chart Consistency**.\\

\#\# Text–Chart Consistency Evaluation Criteria\\

Text–chart consistency refers to how tightly the written analysis is anchored to specific charts and tables. When comparing reports, prioritize those where:\\

* **Unified concepts and metrics**: Key terms and indicators are defined once and then used with the same names, units, and thresholds across text, tables, and figures.\\
* **One-to-one text–figure alignment**: Every major conclusion in the text can be directly traced to a specific chart/table with matching time range, variables, and comparison groups.\\
* **Explicit data scope and limits**: Charts clearly mark data ranges, assumptions, and missing or incomplete data, and the text reiterates these limits when interpreting the results.\\
* **Integrative Summary Visuals**: Dashboards or synthesis figures are used to recap key patterns, and concluding text explicitly walks through these visuals to close the loop.\\

\#\# Input\\
I have uploaded the text-image pairs below, grouped by Report ID and corresponding shortname of generation method.

\end{tcolorbox}
\caption{Prompt for evaluating text-chart consistency.}
\label{fig:tc_prompt}
\end{figure*}

\begin{figure*}[h]
\begin{tcolorbox}[colback=gray!5!white,colframe=gray!75!black,title=Prompt of Depth. evaluation]

\#\# Task\\
Your task is to evaluate and rank the quality of text-image pairs across multiple reports based on **Informative Depth**.\\

\#\# Text–Depth Evaluation Criteria\\
Text–depth refers to how effectively the report converts data and visuals into a structured, system-level story rather than a set of isolated comments. When comparing reports, prioritize those where:\\
* **Multi-Dimensional Coverage**: Figures jointly span time, space, magnitude, and energy (or equivalent key dimensions), forming an integrated dashboards or overview tables rather than a single-angle view.\\
* **Deep Quantitative Interpretation**: Text consistently interprets full distributions and key statistics (e.g., typical levels, variability, skew, anomalies) and ties them back to core questions such as whether patterns are increasing or abnormal.\\
* **Closed-Loop Visual–Text Logic**: Figures and prose are organized around a small set of recurring themes, with later sections integrating earlier findings into concise, decision-ready conclusions.\\

\#\# Input\\
I have uploaded the text-image pairs below, grouped by Report ID.

\end{tcolorbox}
\caption{Prompt for evaluating textual informative depth.}
\label{fig:depth_prompt}
\end{figure*}

\begin{figure*}[t]
\begin{tcolorbox}[colback=gray!5!white,colframe=gray!75!black,title=Prompt of Info. evaluation]

\#\# Task\\
Your task is to evaluate and rank the informativeness of multiple reports.\\

\#\# Report Information Richness\\
Report information richness means the report, through tightly integrated visuals and text, addresses the title’s core question across multiple key dimensions rather than listing isolated results. When comparing reports, prioritize those where:\\

* **Multi-Dimensional Visuals**: Charts within the same report jointly cover temporal trends, spatial patterns, and magnitude–energy relationships in a consistent layout, giving each figure and the overall visual set high information density.\\
* **Explanatory Text Backbone**: The prose is organized around the core question, explains key concepts and metrics, and links observations, plausible causes, and conclusions into a coherent narrative that continuously enriches each visual.\\
* **Integrated Synthesis**: A final integrative summary or dashboard pulls together frequency, intensity, and energy into a compact global view, turning detailed analyses into a reusable analytical overview.\\
* **Content Volume \& Distribution**: The report demonstrates depth through substantial **total word count** and maintains a high quantity of **text and images per chapter**, ensuring consistent detailed coverage across all sections.\\

\#\# Input\\
I have uploaded the report content below. They are grouped by Report ID.

\end{tcolorbox}
\caption{Prompt for evaluating report informativeness.}
\label{fig:info_prompt}
\end{figure*}

\begin{figure*}[t]
\begin{tcolorbox}[colback=gray!5!white,colframe=gray!75!black,title=Prompt of Vis Cons. evaluation]

\#\# Task\\
Your task is to evaluate and rank the visualization consistency of multiple reports.\\

\#\# Visualization Consistency\\
Visualization consistency means the report adheres to a rigorous, unified design language suitable for professional publication, ensuring that distinct visual elements across various sections feel like parts of a single, cohesive system. When comparing reports, prioritize those where:\\

* **Unified Design System**: A strict adherence to a specific color palette and font hierarchy is maintained across all figures. The visual identity (e.g., primary and secondary colors) remains unmistakable whether the viewer is looking at a line chart, a bar graph, or a complex density plot.\\
* **Semantic Visual Logic**: Color and style usage is logical and semantic rather than random. For instance, specific colors represent the specific datasets or variables consistently across different chart types, allowing the reader to track a variable intuitively throughout the report without re-learning the legend.\\
* **Standardized Structural Elements**: There is a meticulous uniformity in the treatment of non-data ink, including gridline opacity, axis label formatting, legend placement, and annotation styles.\\
* **Publication-Ready Polish**: The visualizations demonstrate a flawless execution devoid of style clashes, ensuring that even when chart types vary significantly, the overall visual presentation remains polished and professional.\\

\#\# Input\\
I have uploaded the report content below. They are grouped by Report ID.

\end{tcolorbox}
\caption{Prompt for evaluating visualization consistency.}
\label{fig:vis_cons_prompt}
\end{figure*}

\section{Prompts}
\label{sup: prompts}

In this section, we detail the system prompts and task-specific instructions used to implement the~\methodname{} framework. These prompts are designed to orchestrate the collaboration between the Data-Augmented Analysis Agent and the Real-Time Evidence Construction Writer, ensuring effective task decomposition and coherent text-chart interleaved generation.

\subsection{Report Outline Generation}

The outline generation prompt is used by the report writer $W$ to decompose the user's request $R_{user}$ into a structured high-level outline $O$ conditioned on the dataset overview (Eq.~\ref{eqa: outline}). This prompt follows the plan-write paradigm~\cite{dome}, instructing the writer to first plan the report structure at a coarse granularity before generating detailed content (see the complete prompt in Figure~\ref{fig:outline_prompt}). The outline serves as a roadmap that guides subsequent incremental writing while maintaining flexibility for on-the-fly analysis and visualization.

\begin{figure*}[h]
\begin{tcolorbox}[colback=gray!5!white,colframe=gray!75!black,title=Report Outline Generation Prompt]

\#\# Task

Generate a compelling data report outline centered around User Intent.\\

\#\# Task Details

- Generate an outline of the report following a linear narrative structure considering the data summaries.

- A linear narrative structure is defined as the narrative structure that contain a start, a middle, and an end. Think of it as setting the scene, unveiling the adventure, and wrapping up with a satisfying conclusion.

- Each point in the outline should be broken down into smaller subpoints that highlight specific aspects of the data. These may include: significant figures or patterns, noteworthy exceptions or deviations, and comparisons or changes over time. Add instructions for visualizations (e.g., charts) where necessary.

- The data report’s overarching theme should focus on \{user\_intent\}. Make sure this sentiment is consistent throughout the outline.

- Remember, the essence of a compelling data report is not just in the numbers but in how you tell, so inclusion of visualization instruction is of utmost importance.

- Be specific, be clear, and most importantly, be engaging. The generated outline must coherently and logically relate to the attributes of the data. Be as specific as possible.\\

\#\# User Intent

\{user\_intent\}\\

\#\# Output Format

Generate the outline in a single Markdown code block following this structure:

"""\\
\# Data Report Title\\

\#\# <Section Title Aligned with User Intent>

- Point covering specific aspect of the data

- ...\\

\#\# ...

- ...\\
"""

\end{tcolorbox}
\caption{Report outline generation prompt used by the report writer.}
\label{fig:outline_prompt}
\end{figure*}

\subsection{Incremental Report Writing}

The incremental writing prompt guides the report writer $W$ to generate the text-chart interleaved report segment by segment (Eq.~\ref{eqa: writer_output}). At each step $i$, the writer produces text segment $t_i$ based on the history $H_{i-1}$, outline $O$, and data overview $DO$. The prompt instructs the writer to trigger visualization requests by emitting \texttt{<visualization>} tags when analytical evidence or visual support is needed (see the complete prompt in Figure~\ref{fig:writing_prompt}). This design enables real-time data interaction during the writing process, allowing the writer to request on-demand chart generation from the analysis agent as the narrative evolves.

\begin{figure*}[h]
\begin{tcolorbox}[colback=gray!5!white,colframe=gray!75!black,title=Incremental Report Writing Prompt]
\#\# Task

Your task is to generate a comprehensive data analysis report based on the provided Outline.\\

\#\# Task Details

1. **Focus on User Intent**: The report theme must align with the provided user intent: \{user\_intent\}.

2. **Follow the Outline**: The report must strictly adhere to the structure and narrative flow defined in the outline.

3. **Elaborate on Sections**: Flesh out each section of the outline with detailed analysis, insights, and clear, professional prose

4. **Visualization Requests**: When a visualization is needed, generate a brief visualization request enclosed in `<visualization></visualization>` tags based on the data summaries. The `Visualization Requests` means a natural-language plain text that states there is a need for a visualization to support the analysis. The content in the `<visualization></visualization>` tags will be prompt a data analyst agent to create the chart.

5. **Visualization Result**: When the data analyst agent generates the chart, some result in tag `<visualization\_result></visualization\_result>` including the chart image will be appended following the visualization request tag `<visualization></visualization>`.

6. **Continuation Behavior**: After each `<visualization\_result></visualization\_result>` tag, continue the report with 2-4 sentences of analysis, insights, and clear, professional prose.\\

\#\# User Intent

\{user\_intent\}\\

\#\# Outline

\{outline\}\\

\#\# Visualization Format

Write a concise natural-language visualization request inside `<visualization></visualization>`.
Just state the visualization goal and message.

Example:

"""\\
<visualization>

Compare regions by their average monthly sales in 2024 to identify top and under performing regions. Focus on region names and average monthly sales; exclude regions with fewer than 10 records in 2024.

</visualization>

"""\\

\#\# Output Format

Just output the report text without any explanation with Markdown format.
\end{tcolorbox}
\caption{Incremental report writing prompt used by the report writer.}
\label{fig:writing_prompt}
\end{figure*}

\subsection{Visualization specification planning}

The visualization specification planning prompt $P_{ana}$ is used by the data-augmented analysis agent $A'$ to interpret and respond to visualization requests from the writer. As formulated in Eq.~\ref{eq: vis_plan}, given a visualization request $q$ (enclosed in \texttt{<visualization>} tags), $A'$ applies $P_{ana}$ to extract an analysis intent $i$ and produce a detailed visualization specification $vs$ that includes the referenced table subset $D_{ref}$. The prompt (shown in Figure~\ref{fig:vis_spec_prompt}) guides the agent to leverage its dataset-specific knowledge (from the injected data overview) to select relevant variables, determine appropriate chart types and aggregations, and generate contextually grounded visualization specifications that align with the writer's narrative requirements.

\begin{figure*}[h]
\begin{tcolorbox}[colback=gray!5!white,colframe=gray!75!black,title=Visualization Specification Planning Prompt]

You are a professional data analyst and chart designer.\\

\#\# Task\\
- Your task is to analyze the user input and generate a visualization description with necessary information.\\
- The visualization description must be put in <visualization></visualization> tag.\\

\{chart\_style\}\\

\#\# Data tables\\
The data tables provide the raw data used for Visualization.\\

\{summaries\}\\

\{tables\}\\

\#\# Output Format\\
The output visualization description must strictly follow the following yaml format:\\
"""\\
chart\_type: <Chart type>\\
title: <Title of the visualization>\\
data: <All data to be visualized>\\
labels: <Description of the axis labels, legends, and other text labels>\\
"""

\end{tcolorbox}
\caption{The visualization specification planning prompt $P_{ana}$ used by the data-augmented analysis agent $A'$.}
\label{fig:vis_spec_prompt}
\end{figure*}

\subsection{Caption Generation}

The caption generation prompt $P_{caption}$(shown in Figure~\ref{fig:caption_prompt}) is used by the data-augmented analysis agent $A'$ to produce analysis-oriented insights that directly respond to the writer's visualization requests. As formulated in Eq.~\ref{eq: insight_for}, given a generated chart $c$, the original analysis intent $i$, and the writer's query $q$, $A'$ applies $P_{caption}$ to generate a concise, well-grounded caption $s$ that: (1) directly addresses the analytical question posed in $q$, (2) is strictly grounded in the visual evidence presented in $c$, and (3) aligns with the analysis intent $i$.

\begin{figure*}[h]
\begin{tcolorbox}[colback=gray!5!white,colframe=gray!75!black,title=Caption Generation Prompt]

\#\# Task\\
Your task is to generate a concise, and descriptive caption for the provided Picture with title \{title\} and user intent \{user\_intent\}.\\
The Picture provided is a chart generated based on the `Visualization Request`:\\
\{visualization\_request\}\\

You should generate a caption that accurately and clearly based on the `Visualization Request`.\\

\#\# Output Format\\
Just output the caption in plain text without any explanation.

\end{tcolorbox}
\caption{The analysis-oriented caption generation prompt $P_{caption}$ used by the data-augmented analysis agent $A'$.}
\label{fig:caption_prompt}
\end{figure*}

\section{The Details of Visualization Specification}
\label{sup:vis_spe}

The visualization specification ($vs$) serves as a structured intermediate representation that bridges natural language visualization requests and executable plotting code. Formulated by the analysis agent $A'$, $vs$ encodes all necessary parameters for chart generation in a machine-readable yaml format, ensuring precise and reproducible visual output. This structured approach mitigates ambiguity inherent in natural language requests while maintaining flexibility for diverse visualization types.

A complete visualization specification comprises at least four core components: (1) \textbf{chart\_type}: the visualization category (e.g., line chart, bar chart, scatter plot) that determines the rendering paradigm; (2) \textbf{title}: a descriptive chart heading that summarizes the visualized insight; (3) \textbf{data}: the preprocessed dataset extracted from the original tables, formatted as key-value pairs where each entry represents a data point with its corresponding dimensions and metrics; (4) \textbf{labels}: axis labels, annotations, and auxiliary text elements that enhance interpretability. Figure~\ref{fig: visualization_specification1} and Figure~\ref{fig: visualization_specification2} present two concrete examples of visualization specifications in practice.

\begin{figure*}[h]
\begin{tcolorbox}[colback=gray!5!white,colframe=gray!75!black,title=Example 1 of Visualization Specification]
\textbf{Visualization request:}

Visualize the global trend in smoking prevalence from 2000 to 2022 with a line chart titled “Global Smoking Trends: 2000–2022,” showing a clear downward slope. Annotate key data points: 34.3\% in 2000, 26.7\% average, and 21.7\% in 2022. Include a note indicating a ~12.6 percentage point decline over 22 years.

\textbf{Visualization Specification:}

chart\_type: line\_chart\\
title: Global Smoking Trends: 2000–2022\\
data:\\
  - Year: 2000\\
    Share\_of\_adults\_who\_smoke\_or\_use\_tobacco\_\_age\_standardized\_: 34.322533\\
  - Year: 2005\\
    Share\_of\_adults\_who\_smoke\_or\_use\_tobacco\_\_age\_standardized\_: 30.745779\\
...\\
  - Year: 2021\\
    Share\_of\_adults\_who\_smoke\_or\_use\_tobacco\_\_age\_standardized\_: 22.133286\\
  - Year: 2022\\
    Share\_of\_adults\_who\_smoke\_or\_use\_tobacco\_\_age\_standardized\_: 21.748644\\
labels:\\
  x\_axis\_label: Year\\
  y\_axis\_label: Share of Adults Who Smoke or Use Tobacco (Age-Standardized, \%)\\
  annotations:\\
    - text: "34.3\% in 2000"\\
      position: [2000, 34.322533]\\
      color: "\#000000"\\
      font\_size: 10\\
      arrow: true\\
    - text: "Average: 26.7\%"\\
      position: [2012, 26.677333]\\
      color: "\#000000"\\
      font\_size: 10\\
      arrow: true\\
    - text: "21.7\% in 2022"\\
      position: [2022, 21.748644]\\
      color: "\#000000"\\
      font\_size: 10\\
      arrow: true\\
    - text: "~12.6 percentage point decline over 22 years"\\
      position: [2010, 20]\\
      color: "\#000000"\\
      font\_size: 10\\
      alignment: center\\
  legend: null\\

\end{tcolorbox}
\caption{Example of a visualization specification for a line chart.}
\label{fig: visualization_specification1}
\end{figure*}

\begin{figure*}[h]
\begin{tcolorbox}[colback=gray!5!white,colframe=gray!75!black,title=Example 2 of Visualization Specification]
\textbf{Visualization request:}

Create a horizontal bar chart titled “Cigarette Taxation by Country (2022)” showing the top 10 countries with the highest tax shares (e.g., Bulgaria, Bosnia and Herzegovina, Brazil) and bottom 10 with the lowest (e.g., Brunei, Benin, Antigua and Barbuda). Include a vertical dashed line at the global median of ~53.6\%.

\textbf{Visualization Specification:}

chart\_type: horizontal\_bar\_chart\\
title: "Cigarette Taxation by Country (2022)"\\
data:\\
  - Entity: Brazil\\
    Taxes\_as\_a\_share\_of\_cigarette\_price: 80.25\\
  - Entity: Australia\\
    Taxes\_as\_a\_share\_of\_cigarette\_price: 76.98\\
...\\
  - Entity: Barbados\\
    Taxes\_as\_a\_share\_of\_cigarette\_price: 42.99\\
  - Entity: Algeria\\
    Taxes\_as\_a\_share\_of\_cigarette\_price: 44.16\\
labels:\\
  x\_axis: "Taxes as a share of cigarette price (\%)"\\
  y\_axis: "Country"\\
  median\_line: "Global Median (~53.6\%)"\\
  legend: null\\

\end{tcolorbox}
\caption{Example of a visualization specification for a horizontal bar chart.}
\label{fig: visualization_specification2}
\end{figure*}

\section{The Details of visualization Tools $T$}
\label{sup:vis_tool}

To ensure the quality and reliability of generated visualizations in \methodname{}, the visualization tools $T$ implement a robust three-stage refinement mechanism that addresses the key challenges in automated chart generation: code execution failures, suboptimal visual design, and candidate selection. This mechanism combines (1) iterative code generation with retry logic for error recovery, (2) visual quality enhancement through multimodal feedback, and (3) systematic best candidate selection, ensuring that only high-quality charts are integrated into the final reports.

\textbf{Stage 1: Initial Chart Generation.} The system iteratively generates and executes chart code until a valid chart is produced or the maximum retry limit is reached. At each retry attempt, a text-based LLM $M_t$ generates new visualization code, which is then executed in environment $E$. This process is formalized as:
\begin{equation}
    c_0 = M_t(d),
    \label{eqa: code_gen}
\end{equation}
\begin{equation}
    (success, i_0) = E(c_0),
    \label{eqa: code_exec}
\end{equation}
where $c_0$ is the generated code, $d$ is the visualization description, $success_0$ indicates whether execution succeeded, and $i_0$ is the rendered chart image if successful. The retry loop terminates upon the first successful execution ($success = \text{True}$), yielding the initial chart set $C = \{c_0\}$. If all attempts fail, the system reports failure.

\textbf{Stage 2: Visual Refinement.} Once a successful initial chart $c_0$ is obtained, the system initiates visual refinement through a multimodal LLM critic $M_v$. At each refinement iteration $j$ ($1 \leq j \leq N_{\text{refine}}$), the critic first evaluates the rendered chart image:
\begin{equation}
    feedback_j = M_v(i_{j-1}),
    \label{eqa: critic_eval}
\end{equation}
where $feedback_j$ contains detailed textual critique identifying visual quality issues. A text-based LLM actor $M_t$ then refines the previous code by incorporating the critic feedback:
\begin{equation}
    c_{j} = M_t(c_{j-1}, feedback_j, d).
    \label{eqa: code_refine}
\end{equation}
The refined code is executed using Equation~\eqref{eqa: code_exec}, and if successful, the new chart is added to the candidate set: $C \gets C \cup \{c_{j}\}$. This refinement process continues for up to $N_{\text{refine}}$ iterations, preserving all successfully generated versions.

\textbf{Stage 3: Best Chart Selection.} When multiple chart versions exist ($|C| > 1$), the multimodal critic $M_v$ selects the best chart $\tilde{c}$ by comparing all candidates against the original visualization request $r$:
\begin{equation}
    \tilde{c} = \underset{c \in C}{\mathrm{argmax}}\; M_v(\text{``evaluate quality''}, r, c).
    \label{eqa: best_select}
\end{equation}
If only one chart exists, it is directly selected as $\tilde{c} = c_0$. This three-stage mechanism significantly improves chart reliability and visual quality by separating error handling from aesthetic refinement and ensuring the best candidate is selected through systematic comparison.

\section{The Details of Human Evaluation.}
\label{sup: human_eval}

\subsection{Evaluator Configuration and Sampling Strategy}
To ensure robust and unbiased evaluation, we recruited six evaluators with backgrounds in data analysis. All evaluators were provided with training on the evaluation criteria and assessment protocol prior to the evaluation process.

For each model setting and evaluation dimension, we randomly selected three evaluators from the pool of six to assess that specific dimension. This randomized assignment strategy helps mitigate individual bias and ensures diverse perspectives across different evaluation aspects. Since we evaluate six dimensions across two model settings, each evaluator participated in multiple dimensions while no single evaluator assessed all dimensions for any given setting.

We evaluated report generation under two open-source model settings: (1) \textbf{Qwen3-VL-32B-Instruct} as a representative smaller-scale open-source model, and (2) \textbf{Qwen3-235B-A22B-Instruct-2507 \& Qwen2.5-VL-72B-Instruct} as larger-scale open-source models. From each of the three datasets, we randomly sampled 10 generated reports, resulting in a total of 30 reports evaluated per model setting per dimension. This sampling strategy ensures comprehensive coverage across different domains while maintaining manageable evaluation workload.

\subsection{Evaluation Protocol and Questionnaire Design}

The human evaluation protocol was designed to align strictly with our automated evaluation framework. For each evaluation instance, evaluators were presented with reports generated by different methods (our approach and three baselines) for the same input data. To eliminate bias, the reports were anonymized and randomly labeled as Method A, B, C, and D, with the mapping shuffled across different instances.

Evaluators assessed each report along the same six dimensions used in automatic metrics. For each dimension, evaluators were provided with:
\begin{itemize}
    \item \textbf{Dimension definition}: Clear explanation of what the dimension measures
    \item \textbf{Assessment criteria}: Detailed guidelines on how to judge report quality
    \item \textbf{Reference examples}: Sample reports illustrating different quality levels
\end{itemize}

Rather than assigning absolute scores on a fixed scale, evaluators were instructed to \textbf{rank} the candidate reports from \textbf{1 (Best)} to \textbf{4 (Worst)} for each dimension based on the same assessment criteria.

\section{The details of measuring the information richness of the generated report.}
\label{sup: content_extract}

To quantitatively assess the information richness of generated reports, we employ a proposition-level analysis that decomposes report content into fine-grained semantic units. Our approach consists of two main stages: (1) \textbf{Abstractive Proposition Segmentation (APS)} to extract atomic propositions from reports, and (2) \textbf{Information Filtering and Clustering} to categorize propositions by their information value.

\subsection{Abstractive Proposition Segmentation}

We deploy Gemma-APS-7B~\citep{aps} locally to extract propositions from each generated report. For a report with $N$ sentences, the model produces $M$ propositions where typically $M \geq N$, as complex sentences are decomposed into multiple atomic units.

\subsection{Information Filtering and Clustering}

Once propositions are extracted, we apply an information filtering mechanism to categorize them based on their informational value. This process addresses the challenge that not all propositions contribute equally to report quality---some may be redundant, invalid, or overly verbose. We employ a large language model to classify each proposition into three categories: \textbf{Invalid Information}, which includes propositions that are factually incorrect, hallucinated, or not grounded in the source data; \textbf{Duplicate Information}, referring to propositions that repeat semantic content already expressed elsewhere in the report; and \textbf{Simplified Information}, which consists of valid, non-redundant propositions that contribute unique semantic content. By leveraging APS for fine-grained semantic decomposition and LLM-based information filtering for quality-aware categorization, our evaluation framework moves beyond surface-level text statistics to assess the substantive informational value of generated reports.

\section{Case Study}
\label{sup: case study}

To provide a comprehensive understanding of \methodname's capabilities and limitations, we present two case studies generated by our approach under different scenarios.

\subsection{Good Case Example}

Below is a successful case demonstrating \methodname's effectiveness. The report presents a quantity of sophisticated visualizations and achieves tight text-visual coherence where the text acts as a visual guide—for instance, explicitly describing how ``the green bar (female) extends further than the orange bar (male)'' to interpret the regional gap. Furthermore, descriptions directly match the complex visual encoding, such as instructing the reader to observe that the ``blue violin (female) is positioned lower than the red violin (male)'' to confirm the consistency of mortality rates across regions.

\includepdf[pages=-]{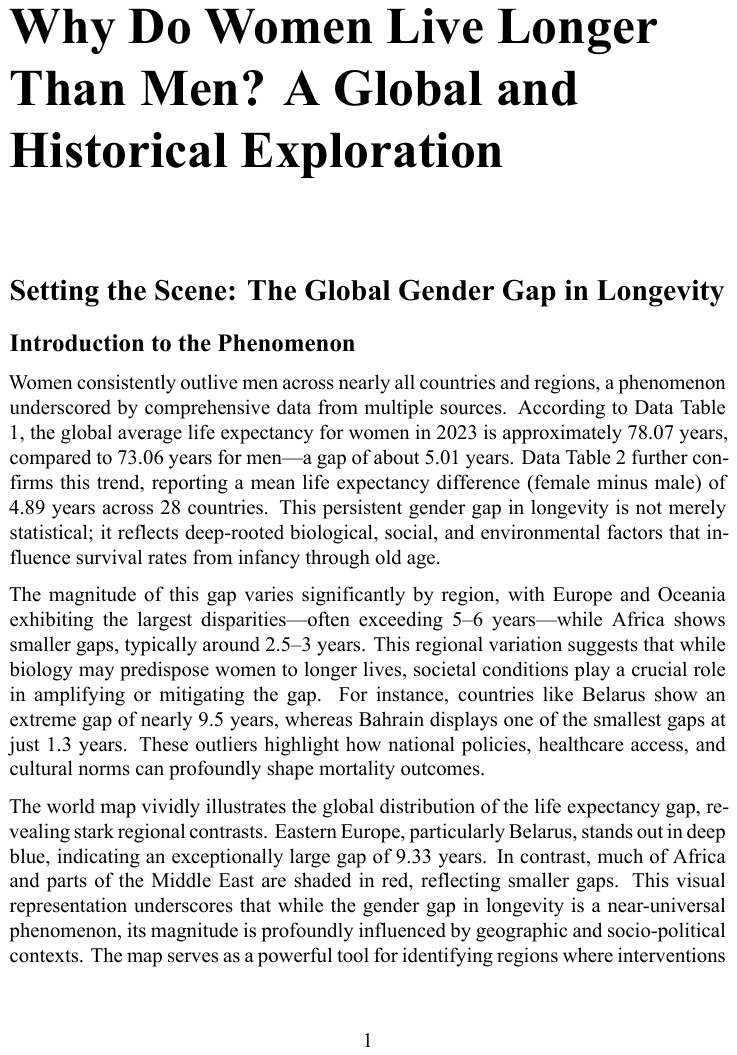}

\subsection{Failure Case Analysis}

Below is a failure case. It reveals a critical cascading failure mechanism in our framework. The report exhibits extremely low visualization density, containing only a basic chart throughout the entire report, which severely limits the analytical depth and decision-support value of the generated content. This low visualization density and quality stems from a deeper technical issue: during the real-time interaction process, the analysis agent $A'$ repeatedly generated visualization code that failed to execute properly due to syntax errors, data type mismatches, or incompatible library function calls. When the code execution failed, $A'$ returned error messages or placeholder outputs instead of valid chart images, which disrupted the text-visual interleaving mechanism. Crucially, these early failures had a compounding negative effect on the writer agent $W$---after experiencing multiple unsuccessful visualization requests where the expected visual evidence was not delivered, $W$ learned to reduce or even abandon its attempts to request new visualizations in later stages of the report generation process. This behavioral adaptation meant that as the report progressed, $W$ increasingly relied on purely textual descriptions rather than leveraging the collaboration framework's core strength of generating text-visual interleaved content, ultimately resulting in a report with minimal visual support and limited analytical value.

\includepdf[pages=-]{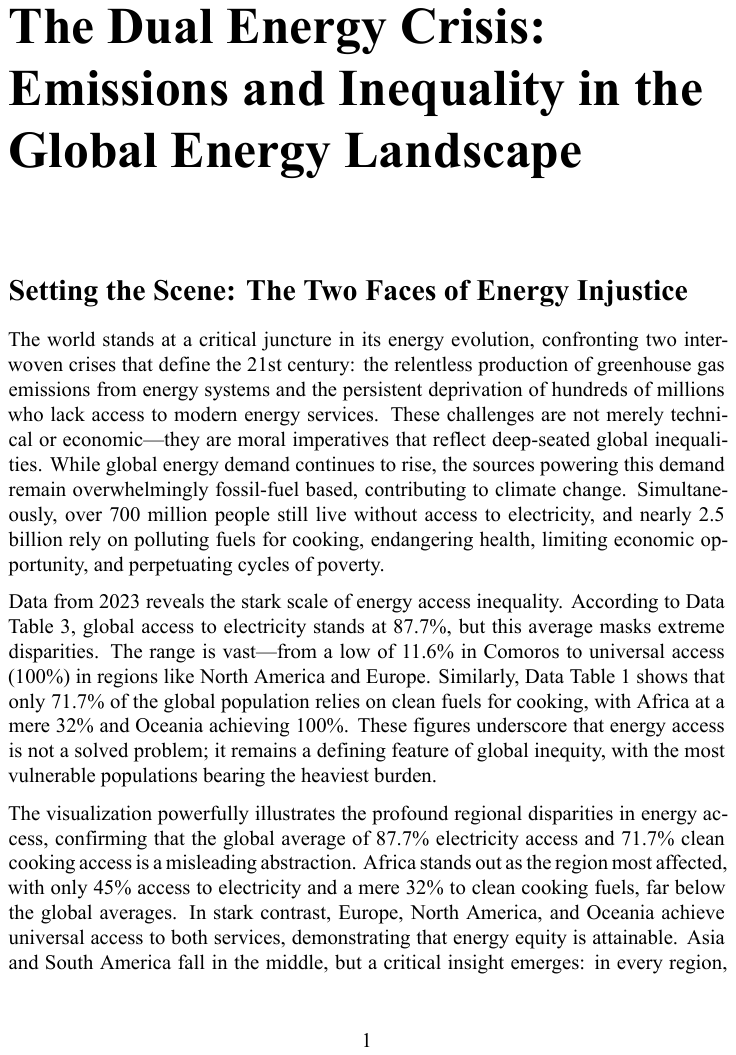}

\end{document}